# Phase-matched Deep Ultraviolet Chiral Bound States in the Continuum Metalens

Omar A. M. Abdelraouf

Institute of Materials Research and Engineering, Agency for Science, Technology, and Research (A*STAR), 2 Fusionopolis Way, #08-03, Innovis, Singapore 138634, Singapore

Correspondence and requests for materials should be addressed to O.A.M.A (email: Omar_Abdelrahman@imre.a-star.edu.sg)

**ABSTRACT**

Coherent light sources in deep ultra-violet (DUV) is essential for various applications such as biomedical imaging and biosensing. Nonlinear crystals (NLCs) generate DUV light, but limited by its bulky thickness, low transparency, and strict narrowband phase-matching conditions. Demonstrating compact, broadband, chiral, and focused DUV light integrated device is elusive. In this work, we present the first demonstration of chiral DUV third harmonic generation (THG) in a dielectric metasurface made from crystalline silicon (c-Si) of compact thickness 400 nm. Metasurface supports chiral bound states in the continuum resonance at fundamental wavelength 800 nm with experimental high $Q$-factor of 130 and modal phase-matched plasmonic resonance at the third harmonic wavelength for efficient THG. Generating DUV-THG power is up to 12 nW using peak power of 15 GW/cm$^2$. Furthermore, we developed a nonlocal metalens operating in DUV for focusing the chiral DUV-THG using the same chiral BIC cavity and phase-gradient approach.  Our platform creates efficient, ultracompact, and multifunctional integrated devices for future integrated DUV nanophotonics devices in MedTech, imaging, and advanced manufacturing.

**KEYWORDs**: dielectric metasurfaces, deep ultraviolet, nonlocal metalens, modal phase-match, chiral bound states in the continuum, third harmonic generation, crystalline silicon, symmetry breaking, and nonlinearity

**INTRODUCTION**

Coherent light sources operating in the deep ultraviolet (DUV, 200–300 nm) spectrum are indispensable for broad applications including photochemistry,[1] nanolithography,[2] spectroscopy,[3] water purification,[4] and high-resolution bioimaging.[5] However, developing compact and efficient DUV sources remains a persistent challenge due to intrinsic material absorption and the scarcity of optical components that perform reliably at these short wavelengths.[6] Traditional DUV light sources such as gas discharge and excimer lasers have been successfully deployed in spectroscopy,[7] micromachining,[8] and surface analysis.[9] Nonetheless, their reliance on bulky instrumentation and gas-handling infrastructure hinders their integration into miniaturized platforms.[10] In contrast, nonlinear optical process called third-harmonic generation (THG), offers an attractive alternative for generating coherent DUV light by upconverting photons from accessible near-infrared (NIR) sources.[11] Since THG does not require non-centrosymmetric media, it can occur in a broad class of materials, including centrosymmetric dielectrics and semiconductors. Despite this, conventional bulk nonlinear crystals suffer from inherent drawbacks such as limited DUV transparency, large physical thickness, and stringent narrowband phase-matching requirements which that constrain their integration and scalability.[11]

Recent progress in dielectric metasurfaces has revolutionized the landscape of optoelectronic devices.[12-26] Subwavelength nanostructured platforms facilitate strong light confinement, high nonlinear response, and ultrathin form factors.[27-32] Compared to their plasmonic counterparts, which suffer from substantial Ohmic losses despite strong field enhancement, dielectric metasurfaces based on materials like crystalline silicon (c-Si) provide a compelling combination of low loss, high damage threshold, and significant third-order nonlinear susceptibility in the NIR regime.[33-36] Furthermore, the plasmonic nature of c-Si near the DUV range enables strong modal overlap at harmonic wavelengths, favoring enhanced nonlinear conversion.[36] Developing high-Q resonances such as bound states in the continuum (BICs) is necessary for boosting light-matter interaction at nanoscale.[37] BICs arise from destructive interference of leaky modes, allowing for deep subwavelength field localization with minimal radiative losses.[38] While theoretical BICs exhibit infinite Q-factors, practical realizations are limited by fabrication imperfections, finite structure size, and coupling efficiency.[39] Even so, BIC-enhanced THG has been successfully demonstrated across the visible and NIR regimes. Extending such mechanisms to the DUV, however, remains underexplored due to increased material absorption and phase mismatch at shorter wavelengths. In particular, the demonstration of chiral BIC-assisted THG in the DUV regime has not been previously reported.

Several recent studies have explored alternative approaches to DUV THG, such as employing transparent conducting oxide material like indium tin oxide (ITO) with plasmonic nanoantennas made from noble metals. In one case, ITO-based metasurfaces were shown to plasmonic resonances near the pump wavelength, enabling modest nonlinear up conversion into the DUV. However, the conversion efficiency was limited due to poor modal overlap and high intrinsic loss in ITO.[40] Likewise, THG from plasmonic metasurfaces composed of gold nanoantennas has been demonstrated, benefiting from field localization near the metal surfaces. Yet these designs suffer from low $Q$-factors, strong Ohmic losses, and the absence of modal phase matching, all of which severely constrain the achievable nonlinear efficiency.[41] Moreover, the spectral bandwidth of enhancement is often narrow and poorly tunable. These limitations underscore the need for metasurfaces that combine high field enhancement, low loss, and carefully engineered mode matching to boost THG efficiency in the DUV regime.

In addition to efficient light generation, shaping and manipulating DUV beams remains a considerable bottleneck. Conventional DUV diffractive optics require ultra-smooth surfaces to suppress scattering, necessitating deep etching and expensive polishing steps that drive up fabrication complexity and cost.[42-44] Moreover, many materials exhibit strong absorption and low refractive index contrast in the DUV, limiting optical performance. The emerging concept of nonlocal metalens addresses these limitations by leveraging long-range electromagnetic coupling between metasurface elements to control far-field harmonic scattering.[45-49] This approach enables precise wavefront shaping at short wavelengths without relying on high-aspect-ratio nanostructures. By integrating generation and focusing into a single ultrathin metasurface, nonlocal metalens holds immense promise for compact and multifunctional DUV nanophotonic platforms, with direct applications in high-resolution imaging, lithography, and biomedical diagnostics.[50-52]

In this work, we overcome these limitations by demonstrating the first chiral BIC-supported THG in the DUV regime using a crystalline silicon metasurface with a compact thickness of 400 nm. Our metasurface supports a high $Q$-factor ($Q \approx 130$) chiral BIC resonance at the fundamental wavelength of 800 nm and a modal phase matched interband plasmonic resonance at the THG wavelength for enabling efficient DUV light generation. The THG power reaches up to 12 nW under an incident peak intensity of 15 GW/cm$^2$, corresponding to a nonlinear conversion efficiency of $3.2 \times 10^{-6}$ %. Importantly, the metasurface incorporates a centro-symmetry broken geometry that facilitates chiral BIC formation and providing handedness selective nonlinear response. Beyond generation, we introduce a novel concept of a nonlocal DUV metalens based on the same chiral BIC cavity. Using a phase gradient design, the metalens enables simultaneous generation and focusing of DUV light by integrating both nonlinear frequency conversion and flat optical manipulation within a single ultracompact device.

**RESULTS AND DISCUSSION**

Figure 1 illustrates the physical mechanism underlying THG from a c-Si metasurface fabricated on a sapphire substrate, designed for DUV nanophotonic light sources. The three-dimensional schematic of the metasurface cavity is presented in Figure 1a. Each unit cell, or meta-atom, is composed of two vertically aligned c-Si nanopillars, each with a height of 400 nm. One pillar is designed as a right-angle trapezoid, while the other retains a rectangular cross-section. A deliberate in-plane symmetry breaking is introduced by varying the difference in width ($\varDelta W$) between the top ($W_2$) and bottom ($W_1$) edges of each pillar. This asymmetry is critical for enabling quasi-BICs with high quality $Q$-factors, both in linear and chiral polarized excitation regimes. Illumination is performed from the top using circularly polarized light, while the THG signal is collected in transmission from the substrate side. Figure 1b presents a scanning electron microscopy (SEM) image of the fabricated chiral c-Si metasurface. The fabrication process began with electron beam lithography (EBL) utilizing hydrogen silsesquioxane (HSQ) resist, followed by pattern transfer into the c-Si layer via inductively coupled plasma reactive ion etching (ICP-RIE). Further details of the fabrication steps are provided in the Methods section and in Figure S1 of the Supporting Information.

Numerical optimization of the metasurface was carried out using finite-difference time-domain (FDTD) simulations. The design parameters such as lattice periodicities ($P_x$ and $P_y$), inter-pillar gap ($g_x$), and bottom pillar width ($W_1$) were initially optimized to tailor the modal interference and achieve efficient coupling into BIC modes. Once these parameters were fixed, the top width ($W_2$) was systematically tuned to maximize the $Q$-factor of the BIC resonance. Figure 1c shows the simulated transmission spectra under linearly polarized excitation, comparing symmetric ($\Delta W = 0$, i.e., $W_1 = W_2$) and asymmetric configurations. In the symmetric case, the structure supports an ideal BIC state fully decoupled from free-space radiation, yielding near-unity transmission. Upon introducing a small asymmetry ($\Delta W = 25$ nm), the BIC couples weakly to the radiation continuum, manifesting as a sharp resonance peak at approximately 800 nm. The asymmetry parameter $\alpha$, defined as $\Delta W/W_1$, plays a pivotal role in tuning the $Q$-factor of the linear BIC resonance.

In Figure 1d, we present simulated transmission spectra under circularly polarized excitation, demonstrating the emergence of chiral BIC resonances due to additional centro-symmetry breaking in the lattice. Two distinct chiral resonances are observed at 800 nm, with the right-handed circularly polarized (RCP) mode achieving a $Q$-factor exceeding 120, while the left-handed circularly polarized (LCP) mode exhibits a lower $Q$-factor of ~70. The differential response between these modes yields a chiral dichroism approaching 50%, signifying strong polarization selectivity in the nonlinear response of the designed chiral BIC cavity.

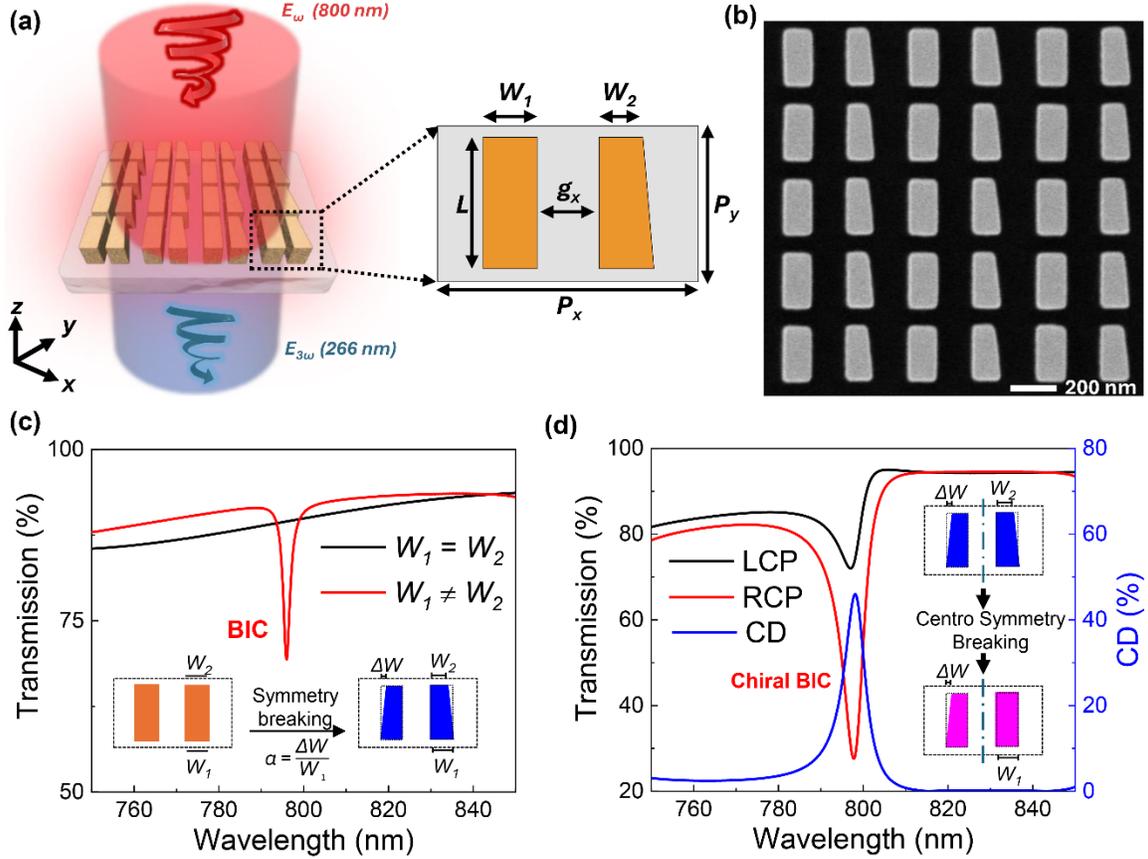

**Figure 1.** Chiral THG generation using chiral BIC cavity within a c-Si metasurface operating in the DUV. (a) Schematic illustration of the c-Si metasurface fabricated on a sapphire substrate. The structure is pumped by a fundamental laser beam ($E_\omega$, $\lambda_1$ = 800 nm), generating a transmitted THG signal ($E_{3\omega}$, $\lambda_2$ = 266 nm) in the DUV. The unit cell design (inset) features fixed geometric parameters: horizontal period ($P_x$ = 500 nm), vertical period ($P_y$ = 290 nm), horizontal gap ($g_x$ = 140 nm), and first side width ($W_1$ = 115 nm). Variable parameters are the vertical gap ($g_y = P_y - L$) and the second side width ($W_2$). The difference $\Delta W = W_1 - W_2$ defines the dimensionless symmetry-breaking parameter $\alpha = \frac{\Delta W}{W_1}$. Pillar height is 400 nm; sapphire substrate thickness is approximately 500 μm. (b) Scanning electron microscope (SEM) image of the fabricated c-Si metasurface on sapphire. Scale bar: 200 nm. (c) Simulated transmission spectra comparing structures with symmetry breaking ($W_1 \neq W_2$, $\Delta W$ = 25 nm) and without symmetry breaking ($W_1 = W_2$). (d) Simulated transmission spectrum demonstrating the effect of centro-symmetry breaking for a single rectangular element. The chiral dichroism (CD) in transmission, defined as $CD = T_{RCP} - T_{LCP}$ (where $T_{RCP}$ and $T_{LCP}$ are the transmission for right- and left-circularly polarized light, respectively), is also shown from simulation.

Multipolar decomposition (MPD) analysis was performed to elucidate the nature of the confined optical modes at the chiral BIC resonance, as illustrated in Figure 2a. The results reveal that the resonant mode primarily arises from the hybridization of an in-plane electric quadrupole (*EQ*) and an out-of-plane magnetic quadrupole (*MQ*), indicating that the chiral BICs originate from higher-order multipolar interference. The simulation procedures for the MPD analysis are described in detail in the Methods section. To explore the dependence of the chiral BIC resonance characteristics on the geometrical asymmetry, we systematically varied the top width $W_2$ of one trapezoidal pillar in meta-atom and computed the corresponding transmission spectra. As shown in Figure 2b, for RCP incident light, the *Q*-factor reaches values as high as 600 when the asymmetry *ΔW* is reduced to 10 nm. The associated transmission minimum becomes significantly deeper, decreasing from 90% to approximately 40%, indicating strong light confinement and enhanced coupling to the far field. In contrast, the transmission response for LCP, shown in Figure 2c, exhibits a more modest resonance dip from 90% to 80%, suggesting weaker coupling and lower modal overlap with the excitation field for this handedness. The corresponding chiral dichroism, defined as the differential transmission between RCP and LCP illumination, is plotted as a function of $W_2$ in Figure 2d. These results demonstrate that the chiral BIC resonance can be finely tuned by adjusting a single geometrical parameter, enabling flexible control over polarization-dependent optical responses across a broad spectral range.

To further understand the field localization properties at the BIC resonance, we computed the spatial distributions of the electric and magnetic field intensities at the resonance wavelength. The magnitude of the electric field ($|E|^2$) and magnetic field ($|H|^2$) are plotted in the horizontal (*x–y*) and vertical (*x–z*) planes through the center height of the pillar ($z$ = 200 nm), as shown in Figure 2e. The electric field is predominantly confined within the horizontal plane, while the magnetic field exhibits strong vertical confinement, consistent with the formation of coupled *EQ* and *MQ* modes. Vector field analysis confirms the characteristic quadrupolar mode profiles responsible for supporting the BIC condition. Moreover, the theoretical upper bound of the *Q*-factor for the proposed chiral BIC structure was assessed as a function of *ΔW*, as depicted in Figure 2f. The simulations predict that by reducing *ΔW* to 1 nm, the *Q*-factor could exceed 50,000, corresponding to an enhancement in localized field intensity by more than two orders of magnitude compared to *ΔW* = 25 nm. This extreme sensitivity to symmetry perturbation underscores the potential of chiral BICs for ultrahigh field confinement and efficient nonlinear light–matter interactions.

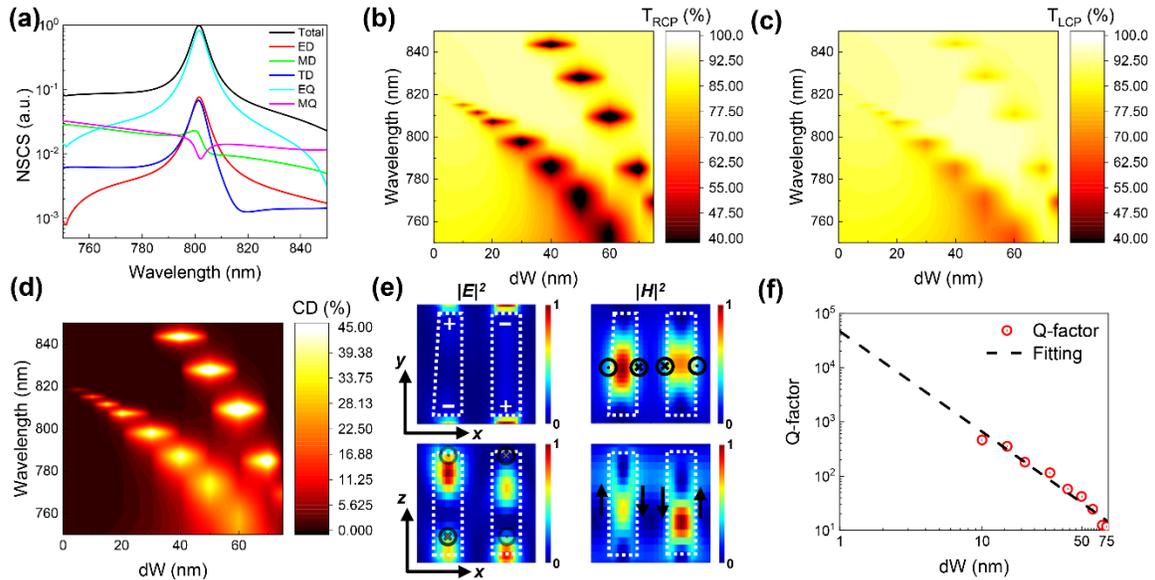

**Figure 2.** Simulated optical and electromagnetic field characteristics of the chiral BIC cavity. (a) Calculated normalized scattering cross section (*NSCS*) spectrum near the chiral BIC wavelength, showing contributions from the electric dipole (*ED*), the magnetic dipole (*MD*), the toroidal dipole (*TD*), the electric quadrupole (*EQ*), and the magnetic quadrupole (*MQ*). (b) Simulated transmission for right-circularly polarized ($T_{RCP}$) light through the chiral BIC cavity as a function of the symmetry-breaking parameter *ΔW*. (c) Simulated transmission for left-circularly polarized ($T_{LCP}$) light through the chiral BIC cavity versus *ΔW*. (d) Simulated chiral dichroism ($CD = T_{RCP} - T_{LCP}$) of the chiral BIC cavity plotted against *ΔW*. (e) Distributions of the vectorial electric field magnitude ($|E|^2$) and magnetic field magnitude ($|H|^2$) within the c-Si metasurface at the chiral BIC resonance wavelength. Profiles are shown in horizontal (*x-y*) and vertical (*x-z*) cross-sections; the horizontal plane is positioned $z = 200$ nm above the pillar base. (f) Simulated quality factor ($Q$-factor) dependence of the chiral BIC metasurface on $\Delta W$.

The experimental verification of the proposed chiral BIC metasurfaces is presented in Figure 3. Series of c-Si metasurface samples were fabricated with varying degrees of asymmetry ($\Delta W$) to investigate their polarization-dependent transmission characteristics and the resulting CD spectra. These measurements aim to demonstrate tunable chiral responses through controlled symmetry breaking, enabling customizable nonlinear chiral optical functionalities. For the symmetric configuration with $\Delta W = 0$ nm, the SEM image in Figure 3a confirms the formation of uniform rectangular meta-atoms, consistent with the design parameters. The measured transmission spectra under LCP and RCP light are shown in Figure 3b. In this configuration, both polarizations exhibit high and featureless transmission, indicating the presence of an ideal BIC that remains completely decoupled from the radiation continuum. This observation is in excellent agreement with the FDTD simulations discussed previously in Figure 2. Upon introducing a small asymmetry of $\Delta W = 10$ nm in Figure 3a, a sharp resonance emerges in the RCP transmission spectrum near 828 nm, signifying the excitation of a chiral BIC mode. In contrast, the LCP spectrum remains unaffected, indicating the polarization selectivity of the resonance. The corresponding measured CD, shown in Figure 3c, exhibits a pronounced peak at the resonance wavelength and a near-zero response outside the resonance band, confirming the selective chiral light–matter interaction induced by the symmetry-broken geometry. When $\Delta W$ is increased to 20 nm, the SEM image in Figure 3a reveals a further deviation from the symmetric geometry due to the reduced trapezoidal pillar volume. This geometric modification results in a blue-shift of the RCP resonance to approximately 823 nm, as shown in Figure 3b, attributed to changes in modal confinement and field distribution within the meta-atoms. The measured $Q$-factor at this condition reaches approximately 117. Although this value is lower than the theoretically predicted $Q$-factor, the reduction is attributed to fabrication-related imperfections such as sidewall roughness and rounded corners, which deviate from the idealized design incorporating sharp interfaces. The corresponding CD spectrum in Figure 3c reflects this blue-shift, maintaining a strong chiral contrast near the modified resonance wavelength. As $\varDelta W$ is further increased from 30 nm to 60 nm, the resonance continues to shift toward shorter wavelengths, reaching approximately 797 nm, as shown in Figure 3b. Simultaneously, the resonance broadens, evidenced by an increase in the full width at half maximum (FWHM), and the $Q$-factor correspondingly decreases. This trend is consistent with the progressive weakening of the symmetry-protected BIC condition due to increasing asymmetry. The overall experimentally observed spectral tuning range of the chiral BIC-induced CD exceeds 30 nm, offering a versatile platform for polarization-selective nonlinear nanophotonic applications.

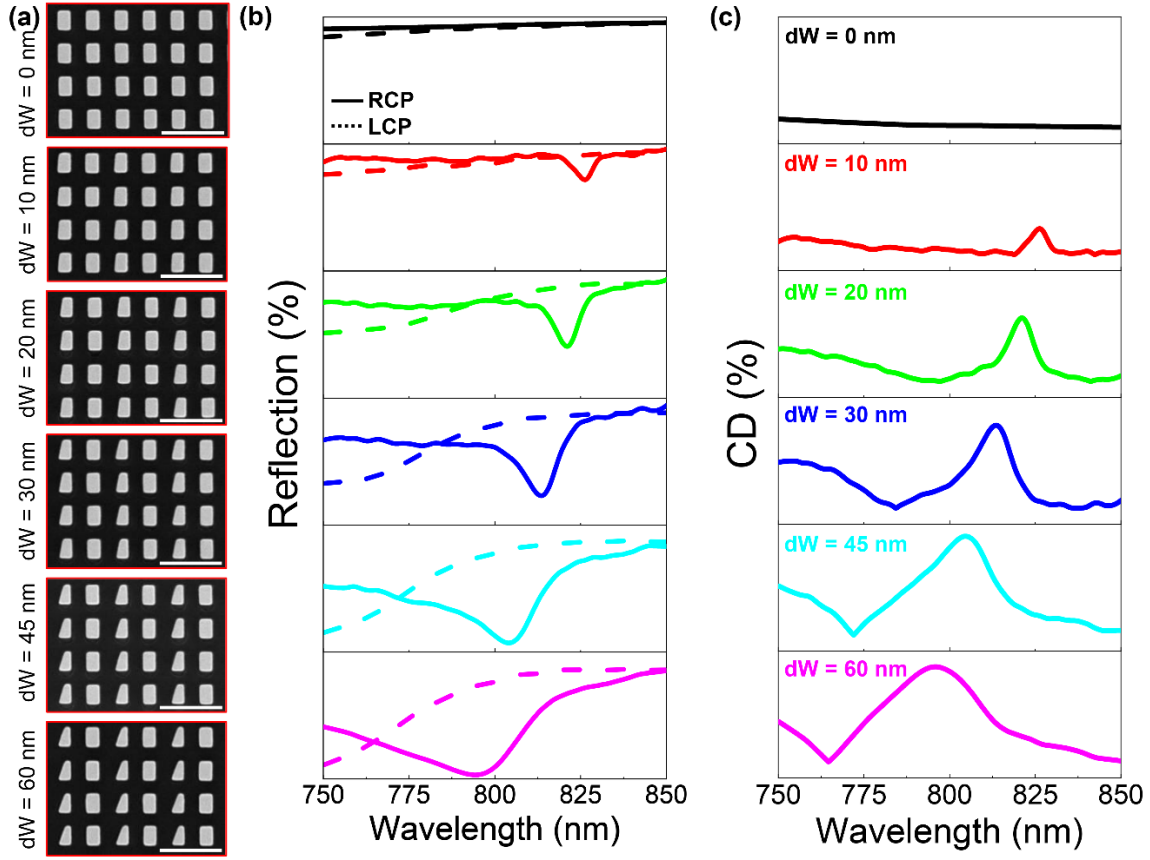

**Figure 3.** Experimental characterization of the chiral BIC cavity performance as a function of the symmetry-breaking parameter $ΔW$. (a) SEM images of the fabricated c-Si metasurfaces showing varying $ΔW$ values. Scale bars: 500 nm (insets). (b) Experimentally measured transmission spectra for right-circularly polarized (*RCP*) and left-circularly polarized (*LCP*) light through the c-Si metasurface across different $ΔW$ values. (c) Measured chiral dichroism of the fabricated c-Si metasurfaces plotted against $ΔW$.

Experimental evaluation of the nonlinear response from the fabricated chiral BIC metasurfaces was carried out using a custom-designed optical setup, illustrated in Figure S3. The RCP THG power in the DUV spectral range was recorded as a function of the incident pump power. As shown in Figure 4a, the metasurface with an asymmetry parameter of $ΔW$ = 25 nm exhibited the highest THG output, reaching a peak power of approximately 12 nW under an average excitation power of 380 mW, corresponding to a fluence of ~1 mJ/cm². This condition reflects optimal coupling between the chiral excitation and the engineered high-$Q$ BIC resonance. THG power measurements for other fabricated asymmetries are summarized in Figure S4, revealing the critical role of symmetry breaking in tuning the nonlinear emission. The experimental data were fitted to the nonlinear power law relationship $P_{THG} = aP_{Pmup}^{b}$, with extracted fitting parameters yielding a slope ~3 and a coefficient of determination $R^2 \approx 0.99$, confirming the cubic dependence expected for a third-order nonlinear process. This scaling indicates efficient phase-matching and resonant enhancement under BIC conditions. The THG conversion efficiency was determined using:

$$\eta = \frac{P_{THG}}{P_{pump}} = 3.2 \times 10^{-6} \,\% \tag{1}$$

While this efficiency is already significant given the low footprint of the metasurface, there remains substantial potential for further improvement. First, enhancing the experimental $Q$-factor by mitigating fabrication-induced imperfections such as sidewall roughness, edge rounding, and etch-induced tapering can narrow the linewidth of the chiral BIC resonance and increase field confinement. Second, the inherent high laser damage threshold of crystalline silicon (~30 mJ/cm² under femtosecond illumination) allows for safe scaling of pump fluence by more than an order of magnitude, offering a path toward at least four orders-of-magnitude enhancement in output THG power without compromising device integrity.[53] To quantify the enhancement due to BIC resonance, a flat c-Si film (400 nm on sapphire) was fabricated as a control structure. Comparison with the metasurface response revealed a remarkable enhancement factor of ~66× at the resonance condition. This enhancement could be further improved by capturing all diffraction orders through a high numerical aperture (NA > 0.85) DUV microscope objective. Supporting evidence for diffraction efficiency and angular distribution is provided in Figure S5.

Spectral dependence of the nonlinear emission was also measured, with the pump wavelength scanned around the near-infrared region. The resulting RCP THG spectrum, plotted in Figure 4b, exhibited a well-defined resonance at 266 nm, corresponding to the third harmonic of 800 nm excitation. Importantly, the measured LCP THG signal was small throughout the scanned range, yielding a measured CD exceeding 80% at the peak resonance. This confirms that the nonlinear emission is not only highly directional and resonant but also intrinsically chiral, due to the selective coupling enabled by the metasurface geometry. Measurements across all fabricated $\varDelta W$ values indicated that increasing asymmetry consistently degraded both the linear and nonlinear responses. As shown in Figure 4c, RCP THG power diminishes with increasing $\varDelta W$, correlating with a reduction in $Q$-factor and modal confinement. While simulations predict stronger THG at lower $\varDelta W$ due to higher Q-factors, a saturation of THG signal was observed experimentally at very small asymmetries. This behavior is attributed to the limited overlapping of the femtosecond laser wavelength emission band and the shorter resonance wavelength supported by BICs with minimal $\varDelta W$, resulting in reduced energy coupling efficiency between laser and metasurface. These results underscore the importance of carefully balancing symmetrical breaking with spectral and pump power alignment to maximize nonlinear output.[54] To further validate the chiral nature of the response, the RCP THG signal was measured as a function of pump polarization, modulated using a quarter-wave plate centered at 800 nm. The data in Figure 4d show a distinct polarization dependence, with maximum THG signal observed under RCP illumination. No appreciable THG was detected under LCP excitation, consistent with the absence of a corresponding LCP BIC mode in the transmission spectrum in Figure 3b. This confirms that the nonlinear optical process is not only polarization-selective but also symmetry-dependent.

Nonlinear numerical simulations based on FDTD methods were performed to examine the spectral behavior and field distributions of the nonlinear response. The RCP and LCP THG intensities were computed using the following third-order polarization integral:

$$P_{THG}(3\omega) = 3\varepsilon_0 \int_V \chi^3_{(3)}(\bar{E}.\bar{E})\bar{E} \qquad (2)$$

where $(\chi_{(3)})$ denotes the third-order nonlinear susceptibility, $(\bar{E})$ is the fundamental electric field. The results, shown in Figure 4e, reveal a sharp THG peak at 266 nm for RCP, while the LCP response remains suppressed to below 10% of the maximum. The simulated CD exceeds 80%, in excellent agreement with experimental observations. Additionally, we performed electromagnetic simulations of the THG power intensity distribution to investigate the influence of geometric design on DUV resonance confinement. Figure 4f presents the THG power profiles at 266 nm, revealing strong localization of the RCP-generated harmonic field with peak amplitude values up to 14. In comparison, the LCP field amplitude remained below 0.3. The resulting RCP/LCP intensity ratio exceeds 46, highlighting the significance of the metasurface in achieving high-purity chiral nonlinear emission. This enhancement originates from inter-band plasmonic resonances in c-Si near the DUV region, which contribute to elevated field confinement at the metasurface–air interface. Additional field mapping plots presented in Figure S2 confirm that these enhancements are most pronounced under RCP excitation, with the electric field $|E|^2$ exhibiting maximal surface-bound localization around the nanopillar edges.

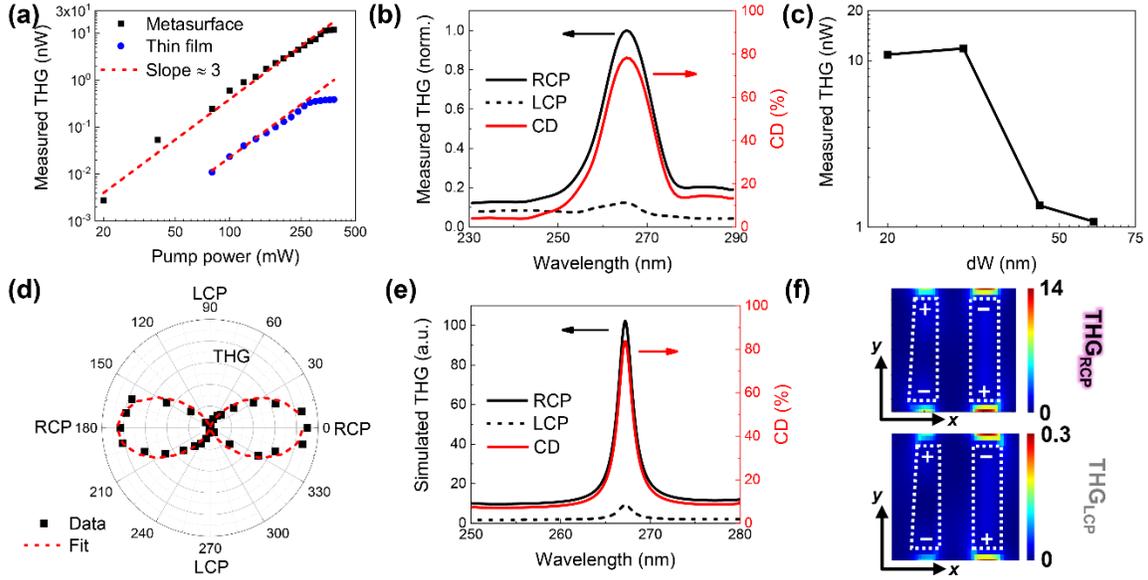

**Figure 4.** Experimental characterization of chiral THG in proposed c-Si metasurface. (a) Measured THG output power as a function of incident pump power for both the RCP chiral BIC metasurface and a reference c-Si thin film. A linear fit to the metasurface data with a slope of ~3 confirms cubic scaling. (b) Measured *THG* emission spectra for right-circularly polarized THG ($THG_{RCP}$), left-circularly polarized THG ($THG_{LCP}$), and the resulting chiral dichroism ($CD_{THG} = THG_{RCP} - THG_{LCP}$). (c) Dependence of measured $THG_{RCP}$ peak power on the symmetry-breaking parameter $\Delta W$. (d) Polar plot showing the measured *THG* power as a function of the rotation angle of a quarter-wave plate (*QWP*) used to generate circular polarization states. (e) Simulated *THG* emission spectra for $THG_{RCP}$, $THG_{LCP}$, and the simulated chiral dichroism $CD_{THG}$. (f) Simulated spatial distribution of *THG* power within a c-Si meta-atom under *RCP* and *LCP* excitation at the chiral BIC resonance, demonstrating a high $THG_{RCP}$ over $THG_{LCP}$ ratio exceeding 46.

We developed a nonlocal DUV metalens that can simultaneously generate chiral THG in DUV and focus the emitted THG in focal point as shown in Figure 5a. The significance of the nonlocal DUV metalens is obtaining on demand compact nanolight source, reducing the need for using large refractive DUV objective and directly shrinking the light generation and focusing for practical integrated applications.[55] The design of nonlocal metalens follows Pancharatnam-Berry parabolic phase profile to focus emitted RCP THG to certain focal point at certain focal length (*f*). More information about phase simulation and metalens phase mapin Figure S6. The fabricated nonlocal metalens is plotted in Figure 5b and it follows same fabrication of chiral BIC in Figure 3a. The metalens focal point at focal plane is illustrated in Figure 5c, the FWHM is 357 nm with corresponding peak power density of 10.8 W/cm$^2$. This is a record peak power density in DUV is sufficient for characterizing materials and biosensing.[5] The far field propagation of metalens along $z$-axis of the nonlocal metalens is plotted in Figure 5d. A clear focused focal point at designed focal distance. The line plot of focal point at focal plane for theorical parabolic lens and metalens are plotted in Figure 5e. The consistent between theory and metalens showed the proper design of nonlocal metalens after optimizing nonlinear phase and optical losses at DUV. We calculated the resolving efficiency of proposed nonlocal metalens in DUV by calculating modulation transfer function as shown in Figure 5f. The proposed nonlocal metalens showed an optical contrast more than 10%, for spatial line frequency of 1000 lines/mm. The degradation in resolving lines in fabricated lens due to residual side loops, which could be improved through optimizing phase map. However, the obtained resolving lines power is sufficient for resolving objects with resolution down to 1 um. However, the experimental demonstration of nonlocal DUV metalens was challenging for the current device, as the THG power reach camera is very small and the EQE efficiency of CCD camera used limit accurate detection. Future works will include improving conversion efficiency of THG power to reach up to 110 uW level using maximum fluence density of c-Si film.[56] In addition, the proposed nonlocal metalens concept could be extended to shorter wavelength emission such as extreme ultraviolet after proper design and material selection. Moreover, incorporating reconfigurable materials instead of c-Si could support varifocal nonlocal metalens for broader range of applications in advanced manufacturing and metrology.

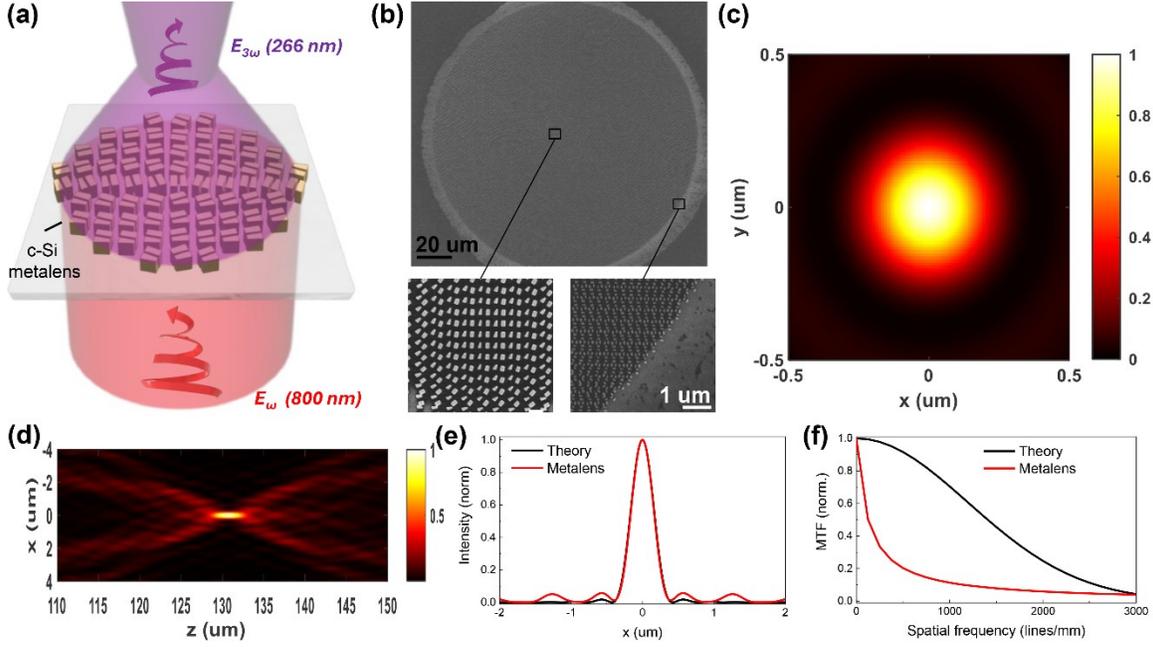

**Figure 5.** Demonstration of a nonlocal DUV-THG chiral metalens based on chiral BIC resonance. (a) Schematic illustration of the nonlocal metalens operating under chiral BIC resonance. The structure is illuminated by fundamental pump light ($E_\omega$, $\lambda_1$ = 800 nm) and generates far-field focused chiral THG emission ($E_{3\omega}$, $\lambda_3$ = 266 nm) in the DUV. (b) SEM images of the fabricated nonlocal chiral BIC metalens. (Inset) Magnified SEM views near the device center and edges. (c) Metalens focused light intensity profiles in the transverse (*x-y*) plane at the metalens focal point. (d) Metalens far-field light intensity distributions in the longitudinal (*x-z*) plane. (e) Theoretical prediction and metalens line plot of the focused light intensity profile along the *x*-axis at the focal plane. (f) Modulation transfer function (MTF) analysis showing the theoretical diffraction-limited performance and metalens MTF curves for the proposed nonlocal DUV chiral metalens.

## CONCLUSION

In conclusion, we have shown the first demonstration in literature for chiral bound states in the continuum operating in the DUV regime made from c-Si metasurface. The experimental chiral BIC high *Q*-factor of 130 and the interband plasmonic resonance at the THG wavelength allowed modal phase-matched response for amplifying THG power up to 12 nW. The nonlinear conversion efficiency is $3.2 \times 10^{-6}$ % using incident peak power density of 15 GW/cm$^2$. Moreover, we designed a nonlocal metalens operating in DUV to focus the chiral DUV-THG. Our results show the promising advantage of nonlinear flat optics technology for achieving multifunctional integrated nanophotonics devices in DUV for bioimaging, MedTech, and advanced manufacturing. Looking ahead, we believe emerging nonlinear 2D materials could enhance the conversion efficiency and allow more light power with thinner material thickness for shorter wavelengths below DUV. Developing and engineering nonlinear metaoptics devices will miniaturize the EUV industry for efficient and accessible EUV lithography, reconfigurable quantum optics,[57-59] and optical computing.[60-62]

**METHODS**

**Transmission and Field Profile Simulations**

Three-dimensional electromagnetic simulations were conducted using a commercial finite-difference time-domain (FDTD) software package (Lumerical FDTD Solution).[63] Material optical properties for crystalline silicon (c-Si) and sapphire were imported directly from the software's built-in database. To probe the chiral bound state in the continuum (BIC) cavity response, excitation was implemented using two orthogonally oriented linearly polarized plane wave sources configured to generate right- and left-circularly polarized (RCP/LCP) illumination. Computational boundaries employed periodic conditions (PBC) along transverse (*x-y*) directions and perfectly matched layers (PML) in the longitudinal (*z*-axis) direction. Spatial discretization featured a base mesh size below 10 nm, with locally refined meshing applied to all c-Si nanostructures. Transmission properties were quantified using a two-dimensional field monitor positioned beneath the sapphire substrate. Field distributions were captured via orthogonal cross-sectional monitors intersecting the BIC cavity center: one horizontal (*x-y* plane) and one vertical (*x-z* plane). All simulations maintained an energy decay termination threshold of <1×10$^{-7}$ to ensure numerical convergence and solution accuracy.

**Multipolar decomposition calculations**

Multipolar decomposition analysis was conducted using Lumerical's FDTD computational engine. Two volumetric monitors centred on the BIC cavity were employed. First, 3D field monitor recorded all electric (*E*) and magnetic (*H*) field vector components within the c-Si metasurface at every mesh point. Second, 3D refractive index monitor mapped the effective refractive index throughout the BIC cavity volume at each mesh location. Scattering cross sections for confined electric and magnetic optical modes were computed according to the following expressions:

$$C_{ED} = \frac{k_0^4}{6\pi\epsilon_0^2 E_0^2} \left| p_{car} + \frac{ik_0}{c}\left(t + \frac{k_0^2}{10}\overline{R_t^2}\right) \right|^2 \quad (2),$$

$$C_{EQ} = \frac{k_0^6}{80\pi\epsilon_0^2 E_0^2} \left| \overline{\overline{Q_e}} + \frac{ik_0}{c}\overline{\overline{Q_t}} \right|^2 \quad (3),$$

$$C_{MD} = \frac{\eta_0^2 k_0^4}{6\pi E_0^2} \left| m_{car} - k_0^2 \overline{R_m^2} \right|^2 \quad (4),$$

$$C_{MQ} = \frac{\eta_0^2 k_0^6}{80\pi E_0^2} \left| \overline{\overline{Q_m}} \right|^2 \quad (5),$$

where $C_{ED}$, $C_{EQ}$, $C_{MD}$, and $C_{MQ}$ denote the scattering cross-section area of the electric dipole, the electric quadrupole, the magnetic dipole, and the magnetic quadrupole, respectively. The scattering power components $p_{car}$, $\overline{\overline{Q_e}}$, $t$, $m_{car}$, and $\overline{\overline{Q_m}}$ denote the electric dipole, electric quadrupole, toroidal dipole, magnetic dipole, and magnetic quadrupole, respectively. The detailed formulas can be found in the literature.[64]

**Fabrication of c-Si BICs metasurface**

Device nanofabrication commenced with dicing a c-Si wafer with 400 nm thickness on a 500 μm sapphire substrate (University Wafers Inc.), followed by sequential rinsing in acetone, isopropyl alcohol (IPA), and acetone. Subsequent ultrasonic cleaning in acetone for 10 minutes was performed, with additional acetone and IPA rinses. Electron beam lithography using an Elionix ELS-7000 involved spin-coating a 6% hydrogen silsesquioxane (HSQ) resist layer at 2000 rpm for 60 seconds, then applying EZspacer at 1500 rpm for 30 seconds to mitigate charging. Excess EZspacer was removed via 10-second nitrogen drying. Metasurface patterning utilized Beamer software with proximity effect correction for dose uniformity and backscatter reduction, employing a 500 pA beam current at 100 kV acceleration voltage with a base dose of 13,000 μC/cm² across a 300×300 μm² field comprising 60,000 exposure points. Development initiated with EZspacer removal through deionized (DI) water rinsing and nitrogen drying, followed by immersion in a 25% NaOH/NaCl (salty developer) for 60 seconds. Development termination occurred via immediate DI water immersion for 60 seconds, succeeded by rinsing in fresh DI water and IPA, concluding with 30-second nitrogen drying. Pattern transfer to c-Si was achieved via inductively coupled plasma reactive ion etching (ICP-RIE; Oxford OIPT Plasmalab) using pure $Cl_2$ gas (22 sccm flow) at 200 W RF power and 400 W ICP power under 5 mTorr chamber pressure at room temperature, yielding a silicon etch rate of ~230 nm/min while retaining a 40 nm HSQ mask residue.

**Transmission measurement**

Transmission measurements used a supercontinuum ns-laser light source (Opera, Leukos laser Inc.) with repetition rate 30 kHz. Quarter-wave plate (QWP) and linear polarizer (LP) were used to pump the chiral BICs cavity with the RCP and LCP light states to excite the chiral BICs resonances. The collected light in transmission was coupled to fiber connected to Ocean Optics (USB4000) spectrometer and a computer.

**THG power and spectrum measurement**

Third harmonic generation power and spectral characterization employed the optical configuration detailed in Fig. S2, utilizing an optical parametric amplifier (OPA) femtosecond laser (Mira 900) operating at 800 nm center wavelength with 200 fs pulse width and 80 MHz repetition rate. The beam diameter was maintained below 1.2 mm, delivering an average peak power of 500 mW. Incident pump power adjustment and chiral THG power sweeps were controlled via a variable neutral density filter. Right- and left-circularly polarized (RCP/LCP) states were generated using a quarter-wave plate (QWP) and linear polarizer (LP). Sample positioning under the laser spot utilizes a glass sheet beamsplitter reflecting white light from an LED. Beam focusing onto the metasurface was achieved with a B-coated plano-convex focusing lens (FL) of 25 mm focal length and 0.5 numerical aperture. Generated DUV power was collected by a high-transmission (>80%) fused silica collimating lens (CL) exhibiting minimal absorption down to 200 nm. Flip mirror M1 directed alignment illumination to a CCD camera, while M2 coupled the fundamental beam to an Ocean Optics USB4000 spectrometer for resonance wavelength coupling at the sample plane. A fused silica prism spectrally separated the DUV harmonics from the fundamental beam with low loss. Mirror M3, a vacuum ultraviolet (VUV)-optimized aluminum coating, reflected DUV harmonics at >90% efficiency. Stray light suppression employed specific filters: two UV hot mirrors, two FF01-300/80-25, one FF01-439/154-25, one FF01-403/95-25, one EO-11-991, and a long-pass filter (FGL435) blocking residual DUV. A Thorlabs PMTSS photomultiplier tube (PMT) converted DUV power to current. THG power quantification involved: immediate PMT current measurement post-metasurface excitation to preclude thermal effects; subtraction of PMT dark current and background scattered pump light contributions; conversion of corrected current to power using the PMTSS radiant sensitivity at the DUV wavelength; and compensation for cumulative optical losses through division by the measured transmission amplitudes of all intervening components (mirrors, filters, lenses, prism).

## ASSOCIATED CONTENT

## SUPPORTING INFORMATION

Figure S1 plots a 3D schematic shows the nanofabrication flow for the chiral BIC cavity. Figure S2 shows the simulated DUV field confinement at the interband plasmonic resonance. Figure S3 plots the nonlinear optical setup used for DUV-THG measurement. Figure S4 illustrates the measured DUV-THG power chiral BIC versus the asymmetric parameter $\varDelta W$. Figure S5 shows the transmitted far-field diffraction of the chiral BICs metasurface at BIC resonance wavelength. Figure S6 shows the nonlocal metalens simulations. A summary table to benchmark nonlocal metasurface for wavelength below 300 nm in literature.

## CORRESPONDING AUTHORS

Email: Omar_Abdelrahman@imre.a-star.edu.sg

## ORCID

Omar A. M. Abdelraouf: https://orcid.org/0000-0002-9065-7414

**AUTHOR CONTRIBUTIONS**

OAMA conceived the idea and the chiral BICs cavity concept and design, carried out FDTD simulations, materials characterization, nanofabrication of the chiral BICs cavity, linear and nonlinear optical measurements, built the THG optical setup, conceived nonlocal metalens concept, and wrote the manuscript.

**Competing Financial Interests**

The authors declare no competing financial interest.

**ACKNOWLEDGMENT**

Authors acknowledge funding from A*STAR under its Career Development Fund (CDF) grant no. C233312016 and SERC Central Research Funds (CRF). Also, the support from Singapore international graduate award (SINGA).

# REFERENCES

(1) Ozaki, Y.; Saito, Y.; Kawata, S. Introduction to FUV and DUV Spectroscopy. In *Far-and Deep-Ultraviolet Spectroscopy*, Springer, 2015; pp 1-16.
(2) Seisyan, R. Nanolithography in microelectronics: A review. *Technical Physics* **2011**, *56* (8).
(3) Kumamoto, Y. Deep-ultraviolet microscopy and microspectroscopy. In *Far-and Deep-Ultraviolet Spectroscopy*, Springer, 2015; pp 123-144.
(4) Matsumoto, T.; Tatsuno, I.; Hasegawa, T. Instantaneous water purification by deep ultraviolet light in water waveguide: Escherichia coli bacteria disinfection. *Water* **2019**, *11* (5), 968.
(5) Kumamoto, Y.; Taguchi, A.; Kawata, S. Deep‐Ultraviolet Biomolecular Imaging and Analysis. *Advanced Optical Materials* **2019**, *7* (5), 1801099.
(6) Ehrt, D. Deep-UV materials. *Advanced Optical Technologies* **2018**, *7* (4), 225-242.
(7) Smith, S. E.; Shanyfelt, L. M.; Buchanan, K. D.; Hahn, D. W. Differential laser-induced perturbation spectroscopy using a deep-ultraviolet excimer laser. *Optics letters* **2011**, *36* (11), 2116-2118.
(8) Pol, V.; Bennewitz, J. H.; Escher, G. C.; Feldman, M.; Firtion, V. A.; Jewell, T. E.; Wilcomb, B. E.; Clemens, J. T. Excimer laser-based lithography: a deep ultraviolet wafer stepper. In *Optical microlithography V*, 1986; SPIE: Vol. 633, pp 6-16.
(9) Pisonero, J.; Fernández, B.; Pereiro, R.; Bordel, N.; Sanz-Medel, A. Glow-discharge spectrometry for direct analysis of thin and ultra-thin solid films. *TrAC Trends in Analytical Chemistry* **2006**, *25* (1), 11-18.
(10) Paetzel, R. Comparison excimer laser–Solid state laser. In *International Congress on Applications of Lasers & Electro-Optics*, 2002; Laser Institute of America: Vol. 2002, p 163015.
(11) Xuan, H.; Igarashi, H.; Ito, S.; Qu, C.; Zhao, Z.; Kobayashi, Y. High-power, solid-state, deep ultraviolet laser generation. *Applied sciences* **2018**, *8* (2), 233.
(12) Abdelraouf, O. A.; Shaker, A.; Allam, N. K. Novel design of plasmonic and dielectric antireflection coatings to enhance the efficiency of perovskite solar cells. *Solar Energy* **2018**, *174*, 803-814.
(13) Abdelraouf, O. A.; Allam, N. K. Towards nanostructured perovskite solar cells with enhanced efficiency: Coupled optical and electrical modeling. *Solar Energy* **2016**, *137*, 364-370.
(14) Abdelraouf, O. A.; Shaker, A.; Allam, N. K. Front dielectric and back plasmonic wire grating for efficient light trapping in perovskite solar cells. *Optical materials* **2018**, *86*, 311-317.
(15) Abdelraouf, O. A.; Allam, N. K. Nanostructuring for enhanced absorption and carrier collection in CZTS-based solar cells: coupled optical and electrical modeling. *Optical Materials* **2016**, *54*, 84-88.
(16) Khodair, D.; Saeed, A.; Shaker, A.; Abouelatta, M.; Abdelraouf, O. A.; EL-Rabaie, S. A review on tandem solar cells based on Perovskite/Si: 2-T versus 4-T configurations. *Solar Energy* **2025**, *300*, 113815.

(17) Atef, N.; Emara, S. S.; Eissa, D. S.; El‐Sayed, A.; Abdelraouf, O. A.; Allam, N. K. Well‐dispersed Au nanoparticles prepared via magnetron sputtering on TiO2 nanotubes with unprecedentedly high activity for water splitting. *Electrochemical Science Advances* **2021**, *1* (1), e2000004.

(18) Siavash Moakhar, R.; Gholipour, S.; Masudy‐Panah, S.; Seza, A.; Mehdikhani, A.; Riahi‐Noori, N.; Tafazoli, S.; Timasi, N.; Lim, Y. F.; Saliba, M. Recent advances in plasmonic perovskite solar cells. *Advanced science* **2020**, *7* (13), 1902448.

(19) Abdelraouf, O. A.; Abdelrahaman, M. I.; Allam, N. K. Plasmonic scattering nanostructures for efficient light trapping in flat czts solar cells. In *Metamaterials XI*, 2017; SPIE: Vol. 10227, pp 90-98.

(20) Abdelraouf, O. A.; Ali, H. A.; Allam, N. K. Optimizing absorption and scattering cross section of metal nanostructures for enhancing light coupling inside perovskite solar cells. In *2017 Conference on Lasers and Electro-Optics Europe & European Quantum Electronics Conference (CLEO/Europe-EQEC)*, 2017; IEEE: pp 1-1.

(21) Abdelraouf, O. A.; Shaker, A.; Allam, N. K. Plasmonic nanoscatter antireflective coating for efficient CZTS solar cells. In *Photonics for Solar Energy Systems VII*, 2018; SPIE: Vol. 10688, pp 15-23.

(22) Abdelraouf, O. A.; Shaker, A.; Allam, N. K. Design of optimum back contact plasmonic nanostructures for enhancing light coupling in CZTS solar cells. In *Photonics for Solar Energy Systems VII*, 2018; SPIE: Vol. 10688, pp 33-41.

(23) Abdelraouf, O. A.; Shaker, A.; Allam, N. K. Design methodology for selecting optimum plasmonic scattering nanostructures inside CZTS solar cells. In *Photonics for Solar Energy Systems VII*, 2018; SPIE: Vol. 10688, pp 24-32.

(24) Abdelraouf, O. A.; Shaker, A.; Allam, N. K. Enhancing light absorption inside CZTS solar cells using plasmonic and dielectric wire grating metasurface. In *Metamaterials XI*, 2018; SPIE: Vol. 10671, pp 165-174.

(25) Abdelraouf, O. A.; Shaker, A.; Allam, N. K. All dielectric and plasmonic cross-grating metasurface for efficient perovskite solar cells. In *Metamaterials Xi*, 2018; SPIE: Vol. 10671, pp 104-112.

(26) Abdelraouf, O. A.; Shaker, A.; Allam, N. K. Using all dielectric and plasmonic cross grating metasurface for enhancing efficiency of CZTS solar cells. In *Nanophotonics VII*, 2018; SPIE: Vol. 10672, pp 246-255.

(27) Halawa, O. M.; Ahmed, E.; Abdelrazek, M. M.; Nagy, Y. M.; Abdelraouf, O. A. Illuminating the Future: Nanophotonics for Future Green Technologies, Precision Healthcare, and Optical Computing. *arXiv preprint arXiv:2507.06587* **2025**.

(28) Abdelraouf, O. A.; Wang, Z.; Liu, H.; Dong, Z.; Wang, Q.; Ye, M.; Wang, X. R.; Wang, Q. J.; Liu, H. Recent advances in tunable metasurfaces: materials, design, and applications. *ACS nano* **2022**, *16* (9), 13339-13369.

(29) Liu, H.; Wang, H.; Wang, H.; Deng, J.; Ruan, Q.; Zhang, W.; Abdelraouf, O. A.; Ang, N. S. S.; Dong, Z.; Yang, J. K. High-order photonic cavity modes enabled 3D structural colors. *ACS nano* **2022**, *16* (5), 8244-8252.

(30) Abdelraouf, O. A.; Wang, X. C.; Goh Ken, C. H.; Lim Nelson, C. B.; Ng, S. K.; Wang, W. D.; Renshaw Wang, X.; Wang, Q. J.; Liu, H. All‐Optical Switching of Structural Color with a Fabry–Pérot Cavity. *Advanced Photonics Research* **2023**, *4* (11), 2300209.

(31) Jana, S.; Sreekanth, K. V.; Abdelraouf, O. A.; Lin, R.; Liu, H.; Teng, J.; Singh, R. Aperiodic Bragg reflectors for tunable high-purity structural color based on phase change material. *Nano Letters* **2024**, *24* (13), 3922-3929.

(32) Abdelraouf, O. A. Broadband Tunable Deep-UV Emission from AI-Optimized Nonlinear Metasurface Architectures. *arXiv preprint arXiv:2506.10442* **2025**.

(33) Chen, S.; Rahmani, M.; Li, K. F.; Miroshnichenko, A.; Zentgraf, T.; Li, G.; Neshev, D.; Zhang, S. Third harmonic generation enhanced by multipolar interference in complementary silicon metasurfaces. *Acs Photonics* **2018**, *5* (5), 1671-1675.

(34) Abdelraouf, O. A.; Anthur, A. P.; Dong, Z.; Liu, H.; Wang, Q.; Krivitsky, L.; Renshaw Wang, X.; Wang, Q. J.; Liu, H. Multistate tuning of third harmonic generation in fano‐resonant hybrid dielectric metasurfaces. *Advanced Functional Materials* **2021**, *31* (48), 2104627.

(35) Abdelraouf, O. A.; Anthur, A. P.; Liu, H.; Dong, Z.; Wang, Q.; Krivitsky, L.; Wang, X. R.; Wang, Q. J.; Liu, H. Tunable transmissive THG in silicon metasurface enabled by phase change material. In *CLEO: QELS_Fundamental Science*, 2021; Optica Publishing Group: p FTh4K. 3.

(36) Abdelraouf, O. A.; Anthur, A. P.; Wang, X. R.; Wang, Q. J.; Liu, H. Modal phase-matched bound states in the continuum for enhancing third harmonic generation of deep ultraviolet emission. *ACS nano* **2024**, *18* (5), 4388-4397.

(37) Zeng, T.-Y.; Liu, G.-D.; Wang, L.-L.; Lin, Q. Light-matter interactions enhanced by quasi-bound states in the continuum in a graphene-dielectric metasurface. *Optics Express* **2021**, *29* (24), 40177-40186.

(38) Hsu, C. W.; Zhen, B.; Stone, A. D.; Joannopoulos, J. D.; Soljačić, M. Bound states in the continuum. *Nature Reviews Materials* **2016**, *1* (9), 1-13.

(39) Overvig, A. C.; Malek, S. C.; Carter, M. J.; Shrestha, S.; Yu, N. Selection rules for quasibound states in the continuum. *Physical Review B* **2020**, *102* (3), 035434.

(40) Ahmadivand, A.; Semmlinger, M.; Dong, L.; Gerislioglu, B.; Nordlander, P.; Halas, N. J. Toroidal Dipole-Enhanced Third Harmonic Generation of Deep Ultraviolet Light Using Plasmonic Meta-atoms. *Nano Letters* **2019**, *19* (1), 605-611. DOI: 10.1021/acs.nanolett.8b04798.

(41) Shi, L.; Andrade, J. R.; Yi, J.; Marinskas, M.; Reinhardt, C.; Almeida, E.; Morgner, U.; Kovacev, M. Nanoscale broadband deep-ultraviolet light source from plasmonic nanoholes. *ACS photonics* **2019**, *6* (4), 858-863.

(42) Ulrich, W.; Rostalski, H.-J. r.; Hudyma, R. Development of dioptric projection lenses for deep ultraviolet lithography at Carl Zeiss. *Journal of Micro/Nanolithography, MEMS and MOEMS* **2004**, *3* (1), 87-96.

(43) Hung, T.-Y.; Su, C.-S. Fused-silica focusing lens for deep UV laser processing. *Applied optics* **1992**, *31* (22), 4397-4404.

(44) Matsuyama, T.; Tanaka, I.; Ozawa, T.; Nomura, K.; Koyama, T. Improving lens performance through the most recent lens manufacturing process. In *Optical Microlithography XVI*, 2003; SPIE: Vol. 5040, pp 801-810.
(45) Malek, S. C.; Overvig, A. C.; Shrestha, S.; Yu, N. Active nonlocal metasurfaces. *Nanophotonics* **2020**, *10* (1), 655-665.
(46) Overvig, A.; Alù, A. Diffractive nonlocal metasurfaces. *Laser & Photonics Reviews* **2022**, *16* (8), 2100633.
(47) Yao, J.; Fan, Y.; Gao, Y.; Lin, R.; Wang, Z.; Chen, M. K.; Xiao, S.; Tsai, D. P. Nonlocal Huygens' meta-lens for high-quality-factor spin-multiplexing imaging. *Light: Science & Applications* **2025**, *14* (1), 65.
(48) Chen, A.; Monticone, F. Dielectric nonlocal metasurfaces for fully solid-state ultrathin optical systems. *ACS Photonics* **2021**, *8* (5), 1439-1447.
(49) Yao, J.; Lai, F.; Fan, Y.; Wang, Y.; Huang, S.-H.; Leng, B.; Liang, Y.; Lin, R.; Chen, S.; Chen, M. K. Nonlocal meta-lens with Huygens' bound states in the continuum. *Nature communications* **2024**, *15* (1), 6543.
(50) Kwon, H.; Sounas, D.; Cordaro, A.; Polman, A.; Alù, A. Nonlocal metasurfaces for optical signal processing. *Physical review letters* **2018**, *121* (17), 173004.
(51) Kim, M.; Lee, D.; Kim, J.; Rho, J. Nonlocal metasurfaces-enabled analog light localization for imaging and lithography. *Laser & Photonics Reviews* **2024**, *18* (7), 2300718.
(52) Kolkowski, R.; Hakala, T. K.; Shevchenko, A.; Huttunen, M. J. Nonlinear nonlocal metasurfaces. *Applied Physics Letters* **2023**, *122* (16).
(53) Lam, Y.; Tran, D.; Zheng, H.; Murukeshan, V.; Chai, J.; Hardt, D. Surface damage of crystalline silicon by low fluence femtosecond laser pulses. *Surface Review and Letters* **2004**, *11* (02), 217-221.
(54) Shcherbakov, M. R.; Shafirin, P.; Shvets, G. Overcoming the efficiency-bandwidth tradeoff for optical harmonics generation using nonlinear time-variant resonators. *Physical Review A* **2019**, *100* (6), 063847.
(55) Liu, Y.; Wang, B.; Hu, L.; Ji, X.; Zhu, T.; Pan, R.; Yang, H.; Gu, C.; Li, J. Ultraviolet metalens based on nonlinear wavefront manipulation of lithium niobate metasurfaces. *ACS Photonics* **2025**, *12* (4), 1857-1864.
(56) Tran, D.; Zheng, H.; Lam, Y.; Murukeshan, V.; Chai, J.; Hardt, D. E. Femtosecond laser-induced damage morphologies of crystalline silicon by sub-threshold pulses. *Optics and Lasers in engineering* **2005**, *43* (9), 977-986.
(57) Abdelraouf, O. A. M. Electrically tunable photon-pair generation in nanostructured NbOCl2 for quantum communications. *Optics & Laser Technology* **2025**, *192*, 113517. DOI: https://doi.org/10.1016/j.optlastec.2025.113517.
(58) Abdelraouf, O. A.; Wu, M.; Liu, H. Hybrid Metasurfaces Enabling Focused Tunable Amplified Photoluminescence Through Dual Bound States in the Continuum. *Advanced Functional Materials* **2025**, 2505165.
(59) Gu, T.; Kim, H. J.; Rivero-Baleine, C.; Hu, J. Reconfigurable metasurfaces towards commercial success. *Nature Photonics* **2023**, *17* (1), 48-58.

(60) Abdelraouf, O. A.; Mousa, A.; Ragab, M. NanoPhotoNet: AI-Enhanced Design Tool for Reconfigurable and High-Performance Multi-Layer Metasurfaces. *Photonics and Nanostructures-Fundamentals and Applications* **2025**, 101379.

(61) Abdelraouf, O. A. M.; Ahmed, A. M. A.; Eldele, E.; Omar, A. A. From maxwell's equations to artificial intelligence: The evolution of physics-guided AI in nanophotonics and electromagnetics. *Optics & Laser Technology* **2025**, *192*, 113828. DOI: https://doi.org/10.1016/j.optlastec.2025.113828.

(62) Abou-Hamdan, L.; Marinov, E.; Wiecha, P.; del Hougne, P.; Wang, T.; Genevet, P. Programmable metasurfaces for future photonic artificial intelligence. *Nature Reviews Physics* **2025**, 1-17.

(63) *Lumerical Inc.,* https://www.lumerical.com/products/; access date **2020**. (accessed.

(64) Paniagua-Domínguez, R.; Yu, Y. F.; Miroshnichenko, A. E.; Krivitsky, L. A.; Fu, Y. H.; Valuckas, V.; Gonzaga, L.; Toh, Y. T.; Kay, A. Y. S.; Luk'yanchuk, B. Generalized Brewster effect in dielectric metasurfaces. *Nature communications* **2016**, *7* (1), 1-9.